\documentclass[%
 aip,
 jap,
 final,twocolumn,
 amsmath,amssymb,
 superscriptaddress,
]{revtex4}

\usepackage{graphicx}
\usepackage{epstopdf}%
\usepackage{subfig}
\usepackage{bm}

\begin{document}

\preprint{AIP/123-QED}

\title{First-principles studies of Ce-doped RE$_2$M$_2$O$_7$ (RE=Y, La; M = Ti, Zr, Hf): A class of non-scintillators}%
\author{A. Chaudhry}
\affiliation{Lawrence Berkeley National Laboratory, 1 Cyclotron Rd., Berkeley, CA 94720, USA.}
\affiliation{Department of Applied Science, University of California, Davis, CA 95616, USA.}
 \author{A. Canning}%
\affiliation{Lawrence Berkeley National Laboratory, 1 Cyclotron Rd., Berkeley, CA 94720, USA.}
\affiliation{Department of Applied Science, University of California, Davis, CA 95616, USA.}
 \email{acanning@lbl.gov.}
\author{R. Boutchko} 
\affiliation{Lawrence Berkeley National Laboratory, 1 Cyclotron Rd., Berkeley, CA 94720, USA.}
\author{M. J. Weber}
\affiliation{Lawrence Berkeley National Laboratory, 1 Cyclotron Rd., Berkeley, CA 94720, USA.}
\author{N. Gr{\o}nbech-Jensen}%
\affiliation{Lawrence Berkeley National Laboratory, 1 Cyclotron Rd., Berkeley, CA 94720, USA.}
\affiliation{Department of Applied Science, University of California, Davis, CA 95616, USA.}
\author{S. E. Derenzo}
\affiliation{Lawrence Berkeley National Laboratory, 1 Cyclotron Rd., Berkeley, CA 94720, USA.}%

\date{\today}

\begin{abstract}
Lanthanum and yttrium  compounds with composition RE$_2$M$_2$O$_7$ (RE=Y, La; M = Ti, Zr, Hf) have high density and high \textit Z and can 
be doped with Ce onto the La and Y sites. This makes these compounds good candidates for Ce activated scintillator $\gamma$-ray detectors
particularly for the hafnate systems which have a very high density. There is disagreement in the literature concerning La$_2$Hf$_2$O$_7$:Ce
as it has been reported to show both bright as well as no Ce activated luminescence by different experimental groups.
We have performed first-principles electronic structure calculations of these compounds doped with Ce using the pseudopotential
method based on the generalized gradient approximation in density functional theory. The positions of the Ce 4\textit f states
relative to the valence band maximum and the position of the Ce 5\textit d states relative to the conduction band minimum (CBM)
of the host material are determined. We find, unlike Ce activated La and Y compounds where the CBM is typically of La 5\textit d or Y 4\textit d
character, that, in these systems the CBM is predominately of \textit d character on the Ti, Zr, Hf atoms. For all these compounds we also find
that the Ce 5\textit d state lies above the CBM which would prevent any luminescence from the Ce site.
\end{abstract}

\maketitle

\section{\label{sec:intro}Introduction}

Today Ce$^{3+}$ is the most common trivalent lanthanide activator for fast and bright scintillation.\cite{Loef:2001az,Shah:2004id,glodo2008mixed} Excellent energy resolution together with high light yield and fast decay time has stimulated active interest in the development of Ce-doped LaBr$_3$ for application in space missions,\cite{kraft2007development} medical imaging\cite{kuhn2006performance} and radio-isotope identification.\cite{milbrath2007comparison} However, even the efficient rare-earth halide scintillators have undesirable attributes like hygroscopic nature and mechanical fragility which pose challenges to manufacturing\cite{Iltis2006359,higgins2008crystal,churilov2008modeling} and, hence, the quest to find new scintillator materials continues. The possibility of computational modeling to aid scintillator discovery has been the subject of several papers in the past.\cite{derenzo1999prospects,klintenberg2001band,klintenberg2002potential,weber2002inorganic} This is pertinent even more now with the availability of a vast literature of known luminescent materials and powerful computational tools (hardware and algorithms) to study underlying physical phenomena and then employ them for selecting promising candidate materials.  

We have recently developed a systematic approach to select candidate materials for bright Ce scintillation based on a set of theoretical criteria derived from studies of more than 100 crystalline compounds by first-principles electronic structure calculations within the framework of density functional theory (DFT).\cite{canning:2010} Our calculations have been validated for known bright Ce scintillators and non-scintillators (compounds not activated by Ce$^{3+}$ dopant). New Ce-activated scintillator materials have also been reported. The objective of our theoretical work is to select candidate materials for synthesis as well as complement experimental work through simulations of promising scintillator materials.\cite{Derenzo:2010so} Since Ce$^{3+}$ activation has been documented in a variety of host materials,\cite{dorenbos20004fn,dorenbos20005d} the effectiveness of our prescreen selection criteria is further crucially tested by how well we are able to predict non-scintillators.

RE$_2$M$_2$O$_7$ (RE = Y, La; M = Ti, Zr, Hf) compounds have been actively studied for applications in nuclear waste removal,\cite{ewing2004nuclear} catalysis,\cite{hwang2000photocatalytic} as hosts for luminescent centers,\cite{ji20052} and recently in the semiconductor industry.\cite{mereu2005interface} Our interest in these materials is because of their high density and high \textit Z which is a desirable attribute for a scintillator material used in $\gamma$-ray detectors.\cite{derenzo2008design} Ce can be readily incorporated as a dopant in many La and Y based scintillator materials, thus it is interesting to study the electronic structure of these materials doped with Ce.
Another reason to study these compounds from a theoretical standpoint stems from some disagreement in the literature regarding La$_2$Hf$_2$O$_7$:Ce. Borisevich \textit{et al.}\cite{borisevich2003luminescence} reported Ce$^{3+}$ scintillation in La$_2$Hf$_2$O$_7$:Ce powders. But, recent papers show no Ce$^{3+}$ luminescence for La$_2$Hf$_2$O$_7$ when doped with Ce.\cite{ji2005part,cherepy2009scintillators} Ce$^{3+}$ luminescence in La$_2$Hf$_2$O$_7$:Ce is also counter-intuitive from simple empirical electronegativity considerations\cite{weber2010} since electronegativities of La (1.1) and Ce (1.12) are smaller than Hf (1.3) on the Pauling scale. This would imply that Ce 5\textit d character states should lie above Hf 5\textit d states. However, the electronegativity argument does not capture physical considerations such as local structure around the dopant, bonding, crystal field splittings, etc. In particular there are many cases of Ce doped La systems where the Ce 5\textit d states lie above the La 5\textit d states e.g. La$_2$O$_3$:Ce and LaAlO$_3$:Ce, even though the Ce has slightly higher electronegativity than La.\cite{canning:2010} \textit{Ab initio} calculations based on density functional theory have been successfully used by our group\cite{canning2009first,chaudhry2009first,boutchko2009cerium} as well as by others\cite{Andriessen:2007nd,watanabe2008experimental} to study Ce 4\textit f and 5\textit d level locations in known scintillator  and non-scintillator materials. 
In this paper, we present results from first-principles bandstructure calculation of Ce doped RE$_2$M$_2$O$_7$ class of compounds. The relative position of Ce 4\textit f and 5\textit d with respect to the host valence and conduction band edge, are computed to determine the possibility of Ce$^{3+}$ luminescence in these materials.

\section{\label{sec:theory}Theory}

Scintillation in Ce doped materials corresponds to a radiative transition from the Ce excited state ([Xe]4\textit f$^0$5\textit d$^1$), usually referred to as (Ce$^{3+}$)$^*$, to the ground state ([Xe]4\textit f$^1$5\textit d$^0$). Our theoretical calculations for the prediction of candidate scintillator materials are based on the calculation of the Ce 5\textit d and 4\textit f levels relative to the valence and conduction band of the host material. The Ce 4\textit f and 5\textit d levels must lie in the forbidden energy gap of the host material for Ce-activated luminescence to be observed in the material.  

The basic model for scintillation in a Ce doped material is that an incident gamma ray will produce a large number of electron-hole (e-h) pairs in the host material that then transfer to the Ce site. However, when the Ce 4\textit f level is in the valence band (Figure 1(a)) then a hole created by the $\gamma$-ray cannot be captured by the Ce and it may migrate to the top of valence band (VBM). This may result in fluorescence from the host but not Ce$^{3+}$ scintillation. On the other hand, if Ce 5\textit d states are inside the conduction band then the electron created by the $\gamma$-ray will not diffuse to the Ce site since that is not energetically favorable (Figure 1(b)). 
Other possible scenarios for reduced or a lack of detectable Ce$^{3+}$ scintillation arise from loss of e-h pairs to competing processes in the host material\cite{rodnyi1997,Nikl:2008fr} (e.g., hole trapping, electron trapping, STEs) before transfer to the Ce site. 


\begin{figure}[h]
\begin{center}
\subfloat[]{\label{fig:1a}\includegraphics[scale=0.5]{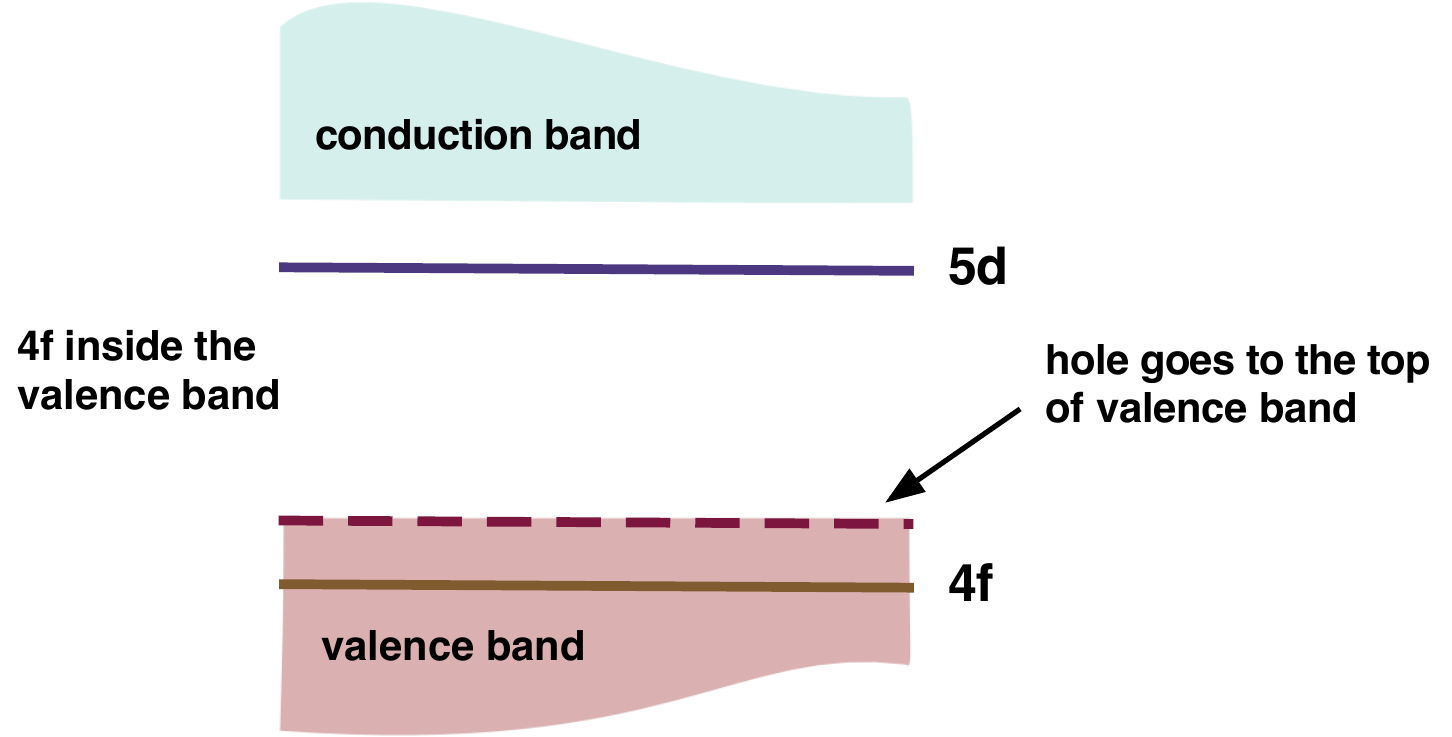}} %
\vspace{6pt}
\subfloat[]{\label{fig:1b}\includegraphics[scale=0.5]{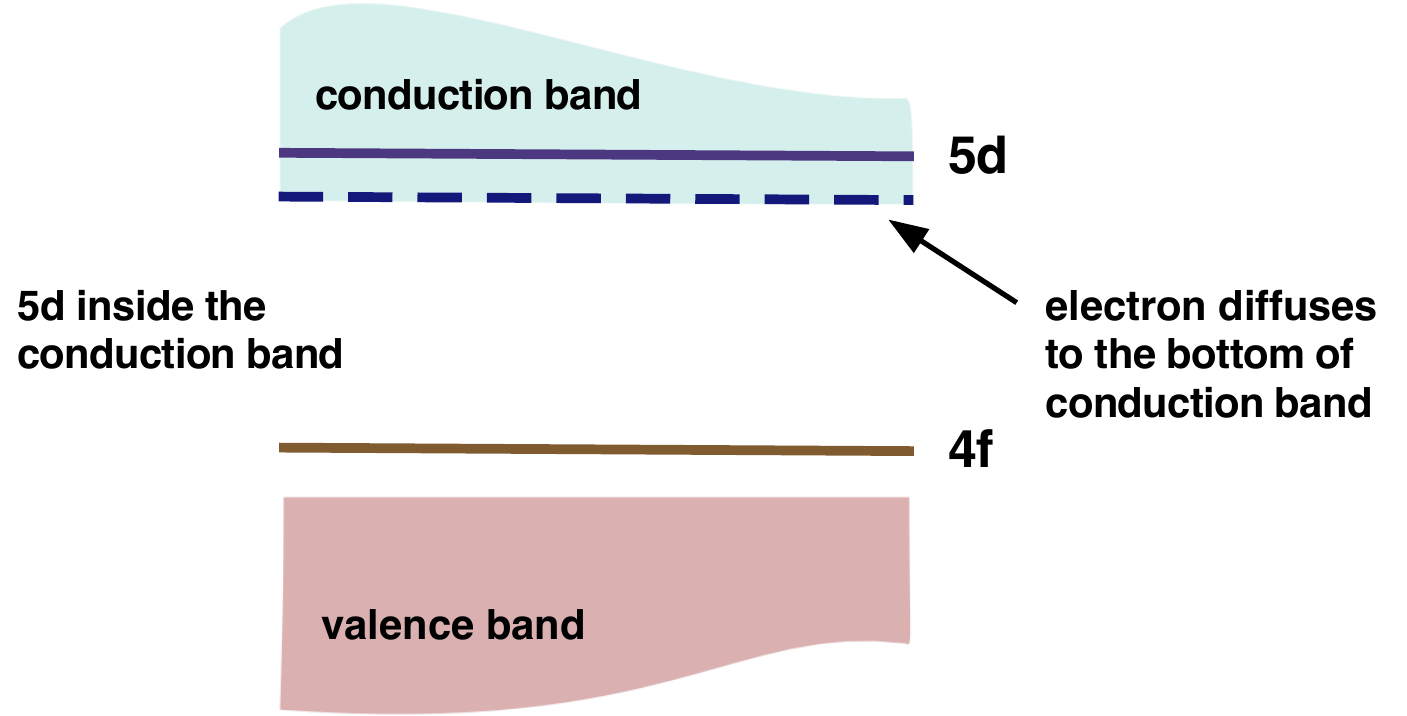}}    
\end{center}
\captionsetup{justification=justified}
\caption{\label{fig:fig1} Simple schematic model to illustrate two possible scenarios for no Ce$^{3+}$ luminescence. \subref{fig:1a} Ce 4\textit f below the top of host valence band \subref{fig:1b} lowest Ce 5\textit d inside the host conduction band.}
\end{figure}

Using first-principles calculations we can predict the Ce 4\textit f-VBM energy gap in a number of host materials, and by studying the character of the lowest occupied conduction band state we can qualitatively estimate if the Ce 5\textit d states are below the conduction band.\cite{canning2009first} We do not model other competing processes in the host which may reduce brightness. However, it should be noted that for Ce$^{3+}$ luminescence to be observed the Ce 4\textit f and 5\textit d states must \textquotedblleft \textit{fit in}\textquotedblright the bandgap of the host. This is a necessary (but not sufficient) condition for Ce$^{3+}$ luminescence.

\section{\label{sec:procedure}Computational Procedure and Details of Calculations}

Our calculation procedure follows the prescription described in Ref.~[\onlinecite{canning:2010}]. We begin by taking atomic positions and crystal symmetry group information of the host compound from the Inorganic Crystal Structure Database (ICSD).\cite{BERGERHOFF:1983zi,ICSD} The calculations are then performed in the following three stages. 

\begin{enumerate}
\item
The host material is doped with a single Ce atom and the atomic positions are relaxed keeping the simulation cell volume and dimensions fixed. We used a large supercell in order to minimize the interaction of the Ce$^{3+}$ dopant with its periodic images. From computational considerations, a supercell matrix containing 88 atoms was chosen for all calculations. By comparing with different cell sizes, this size of cell was found to give reasonably well converged results for the properties of interest. A single Ce$^{3+}$ ion replaces 1 in 16 ions of Y$^{3+}$/La$^{3+}$ in the doped supercell.
\item
A ground state band structure calculation is performed to determine the position of the Ce 4\textit f level with respect to the VBM of the host material. The LDA+U approach\cite{Anisimov:1997io} has been used to give a better description of the localized Ce 4\textit f state.
\item
A constrained LDA (excited state) calculation is then performed by setting the occupancy of Ce 4\textit f states to zero and filling the next highest state. Subsequent analysis of the band decomposed charge density is used to qualitatively determine the position of Ce 5\textit d relative to the bottom of the conduction band (CBM).
\end{enumerate}

All calculations were performed using the plane wave basis set code VASP.\cite{KRESSE:1993kq,Kresse:1996kf,Kresse:1996cu} Spin polarized calculations were carried out within the Perdew-Burke-Ernzerhof (PBE) parameterization to the exchange-correlation functional.\cite{Perdew:1996oq} and the projected augmented wave (PAW) approach,\cite{BLOCHL:1994jv} as implemented in VASP,\cite{Kresse:1999un} was used to treat the valence electron-ion interaction. The plane wave basis set for the electronic wave functions was defined by a cutoff energy of 500 eV. Integration within the Brillouin zone was performed on a $\Gamma$ centered $2\times2\times2$ grid of \textit k-points. The total energy convergence criterion was set to 10$^{-6}$ eV. Structural relaxations of atomic positions were carried out until the residual forces on any atom were less than 0.01 eV/\AA.

We have used the rotationally invariant method of Dudarev \textit{et al.}\cite{Dudarev:1998xi} as implemented in the VASP code\cite{Rohrbach:2003tu} for PBE+U calculations. We have used a value of U$_{\rm eff}$ = 2.5 eV for the Ce 4\textit f states in these compounds where the Ce impurity is surrounded by nearest neighbor oxygen anions. This is based on our first-principles studies of Ce-doped compounds.\cite{canning:2010} The value of U$_{\rm eff}$ = 2.5 eV was found to be suitable for matching the Ce 4\textit f-VBM gap to experiments for a variety of Ce doped compounds including oxides. DFT-PBE (or LDA) calculations incorrectly position the unoccupied La 4\textit f states at the bottom of conduction band as, for example, in LaBr$_3$:Ce.\cite{singh2008applications} However, La 4\textit f states lie higher in energy\cite{czyyk1994local} so following the prescription of Ref.~[\onlinecite{okamoto2006lattice}] we used a U$_{\rm eff}$ parameter for La 4\textit f states and move them higher in energy. An interesting aspect of RE$_2$M$_2$O$_7$ (RE = Y, La; M = Ti, Zr, Hf) compounds is that the transition metal ions have the same +4 valence, but with increasing principal quantum number the \textit d-orbital becomes increasingly extended in character\cite{cox1992} from Ti (3\textit d) to Zr (4\textit d) to Hf (5\textit d) compounds. Ti 3\textit d states are most localized in nature and hence ill described by LDA (or GGA) because of the associated self-interaction error.\cite{solovyev1996t} Our PBE+U calculations include a +U correction for Ti 3\textit d states with U$_{\rm eff}$ = 5.5 eV.\cite{calzado2008effect}

A constrained LDA calculation is performed at the $\Gamma$ point by manually setting the occupation of the Ce 4\textit f states to zero and filling the first excited \textit d character state. From the analysis of the band decomposed charge density we determine the localization of the excited state on the Ce site by measuring the percentage of this state surrounding the Ce atom. This is done by constructing a Voronoi type polyhedron centered on each ion and integrating the portion of charge density within this volume. 

\section{\label{sec:results}Results and Discussion}

\subsection{\label{sec:gs}Ground State}

The La compounds studied in this work mostly crystallize in the cubic pyrochlore structure (space group \textit{Fd3m}) except La$_2$Ti$_2$O$_7$, which exists in a monoclinic layered perovskite structure (space group \textit{P1121}) at room temperature. %
 The pyrochlore structure\cite{subramanian15gv} (A$_2$B$_2$O$_7$) is a super structure derivative of the simple fluorite structure (AO$_2$ $\equiv$ A$_4$O$_8$), where the A and B cations are ordered along the $<$110$>$ direction and one-eigths of the anions are absent. The anion vacancy resides in the tetrahedral intersticial site between adjacent B-site cations which reduces the coordination of cation B from 8 to 6.
 The pyrochlores 
are generally formed by combining a trivalent �A� cation and a tetravalent �B� cation. 
The monoclinic structure is a variant with same stoichiometry as the pyrochlore structure but it is preferred when cation A is larger than cation B. 


\begin{figure}[h]
\begin{center}
\includegraphics[trim=0mm 0mm 0mm 0mm,clip,scale=0.5]{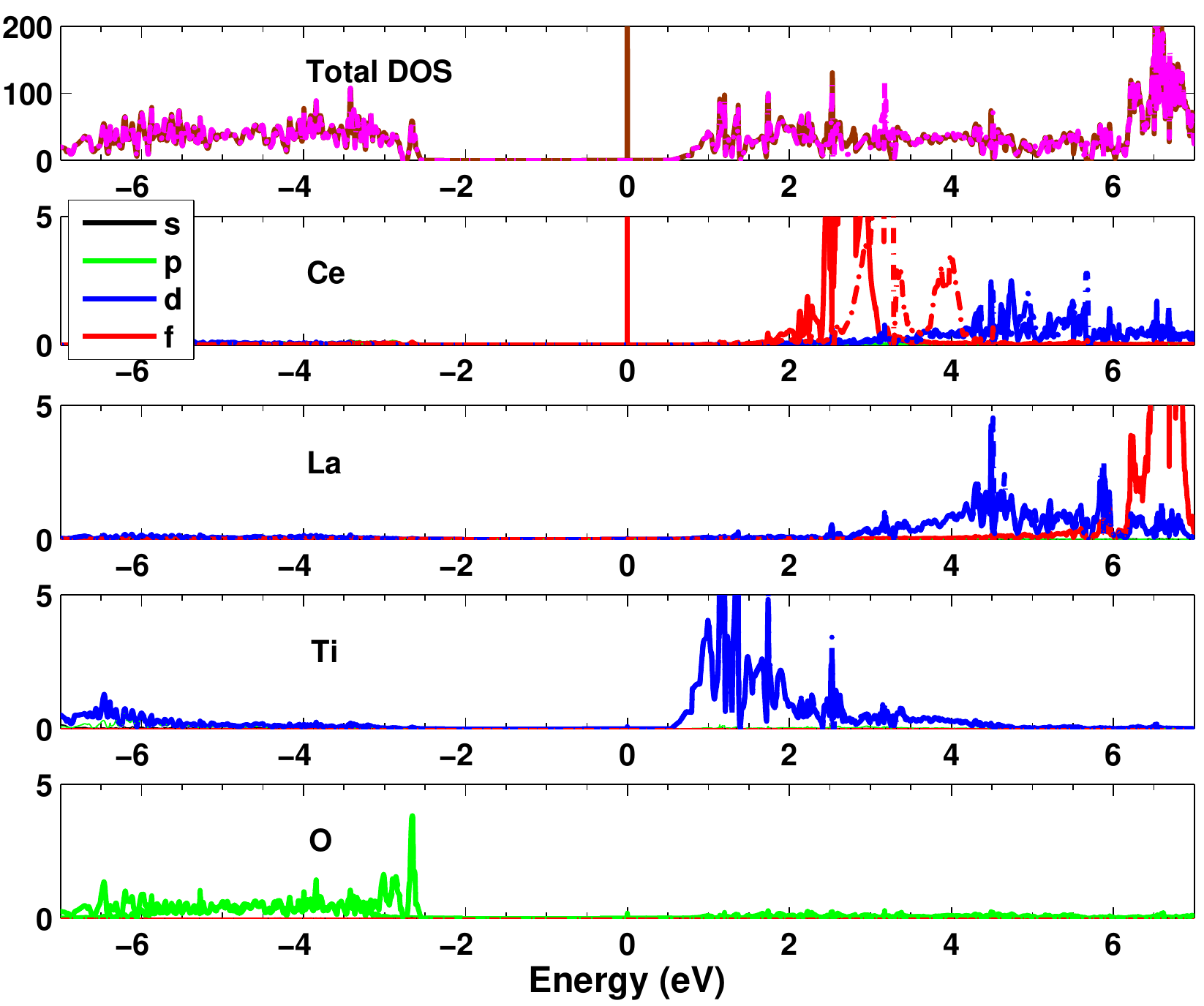}
\end{center}
\caption{\label{fig:fig2} Total and atom projected partial density of states plot for La$_2$Ti$_2$O$_7$:Ce from PBE+U spin polarized calculations (U$_{\rm eff}$ = 2.5eV for Ce 4\textit f states, U$_{\rm eff}$ = 10.36eV for La 4\textit f states and U$_{\rm eff}$ = 5.5eV for Ti 3\textit d states). Spin up states are shown in solid lines and spin down with dashed lines. Ti 3\textit d states are at the bottom of conduction band while Ce 5\textit d states are higher in energy. Oxygen 2\textit p forms the top of valence band.}
\end{figure}

Figure~\ref{fig:fig2} shows the total and atom projected partial density of states (DOS) for monoclinic La$_2$Ti$_2$O$_7$:Ce. The Fermi level is set at 0 eV which is also the position of the occupied Ce 4\textit f state. Oxygen 2\textit p states mainly form the top of valence band and the host VBM is separated from the occupied Ce 4\textit f (i.e., Ce 4\textit f-VBM gap) by approximately 2.5 eV. We can see clearly from the ground state DOS plots that the conduction band minimum (CBM) has a predominantly 3\textit d character attributed to Ti. Ce (and La) 5\textit d states are about 2-3 eV higher in energy. Therefore, it is highly unlikely that an ionized electron will localize on the Ce site. An important point to note here is that the nature of the CBM is markedly different from known oxide Ce scintillators such as LaB$_3$O$_6$:Ce and YAlO$_3$:Ce, where La or Y \textit d character states are generally the lowest lying host conduction band states. 

\begin{figure}[h]
\begin{center}
\includegraphics[trim=0mm 0mm 0mm 0mm,clip,scale=0.5]{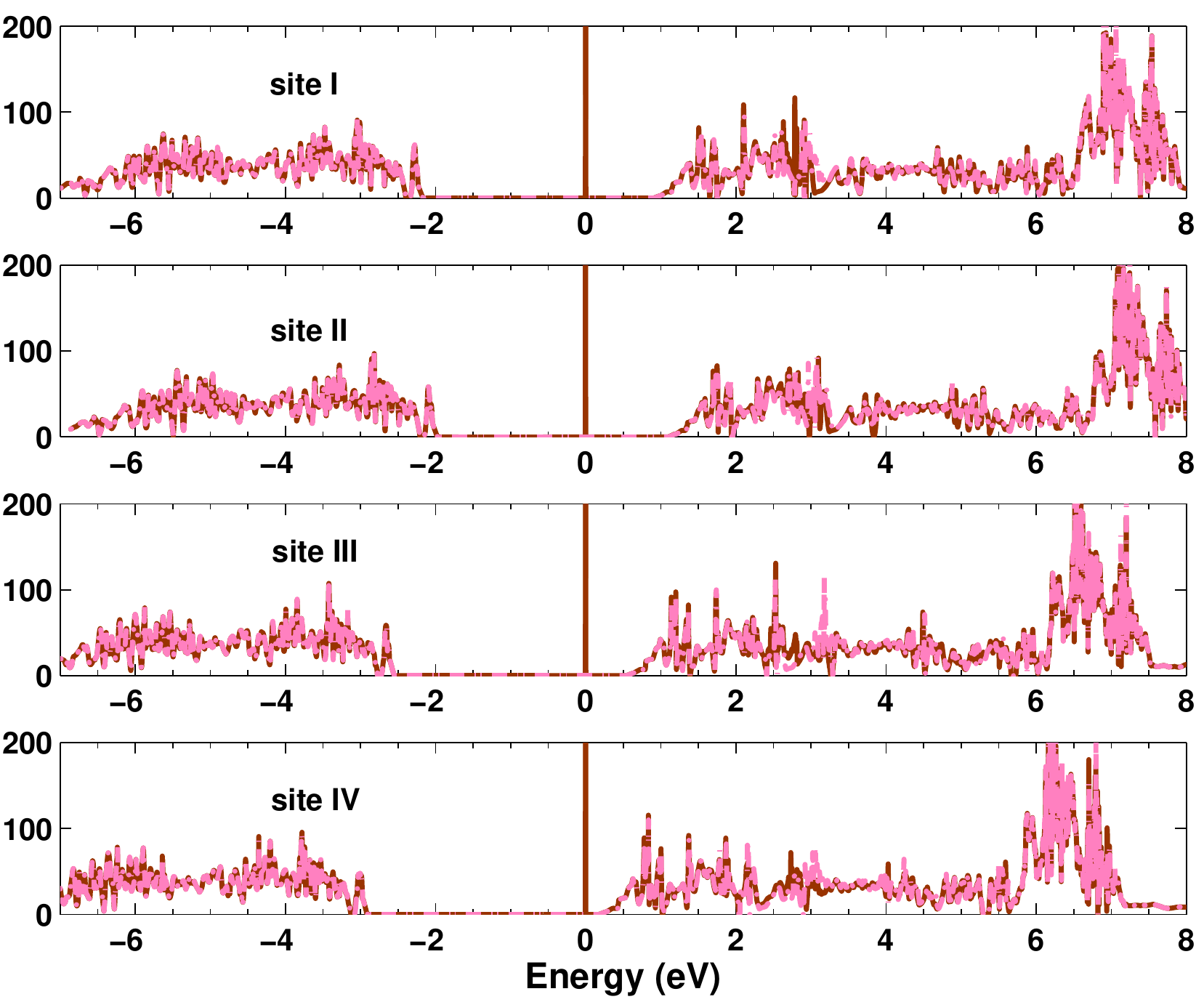}
\end{center}
\caption{\label{fig:fig3} Total density of states plot for La$_2$Ti$_2$O$_7$:Ce for different Ce substitution sites from PBE+U spin polarized calculations. Spin up states are shown in solid lines (brown) and spin down with dashed lines (magenta).} 
\end{figure}

Amongst all the compounds studied in this work, only La$_2$Ti$_2$O$_7$ has more than one inequivalent trivalent sites (La$^{3+}$) for Ce$^{3+}$ substitution. These La sites differ in their oxygen coordination and bond lengths. We found that the Ce 4\textit f-VBM gap varies between 2 eV to 2.9 eV for the different Ce$^{3+}$ substitution sites (Figure~\ref{fig:fig3}). However, since the Ce 5\textit d is situated well above CBM, even though the crystal field effects change due to change in local environment, Ti 3\textit d states are still at the bottom of conduction band. Figure~\ref{fig:fig2} shows partial atom projected DOS plots when Ce is substituted on La site III.


\begin{figure}[h]
\begin{center}
\subfloat[]{\label{fig:4a}\includegraphics[scale=0.5]{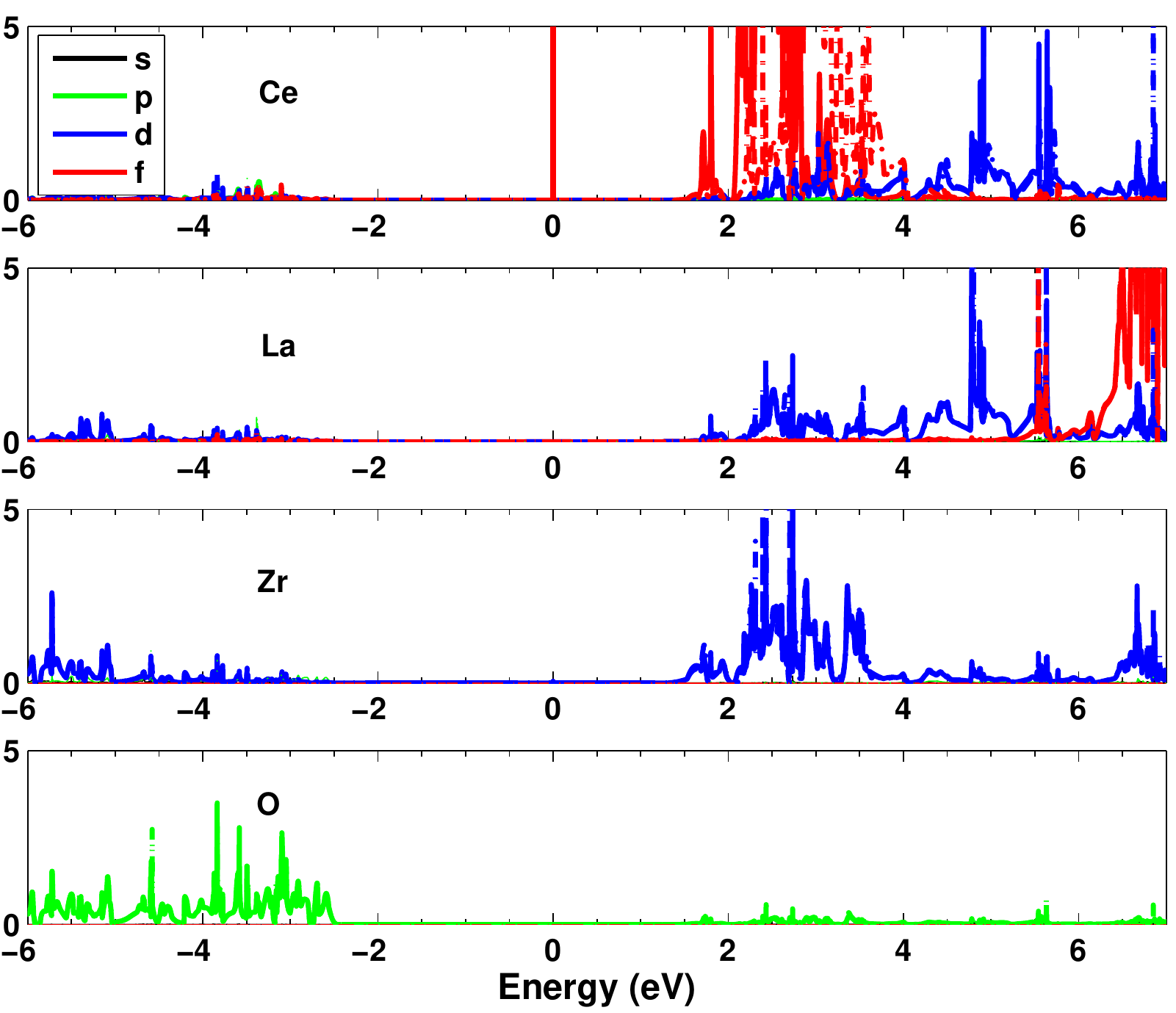}} \qquad
\subfloat[]{\label{fig:4b}\includegraphics[scale=0.5]{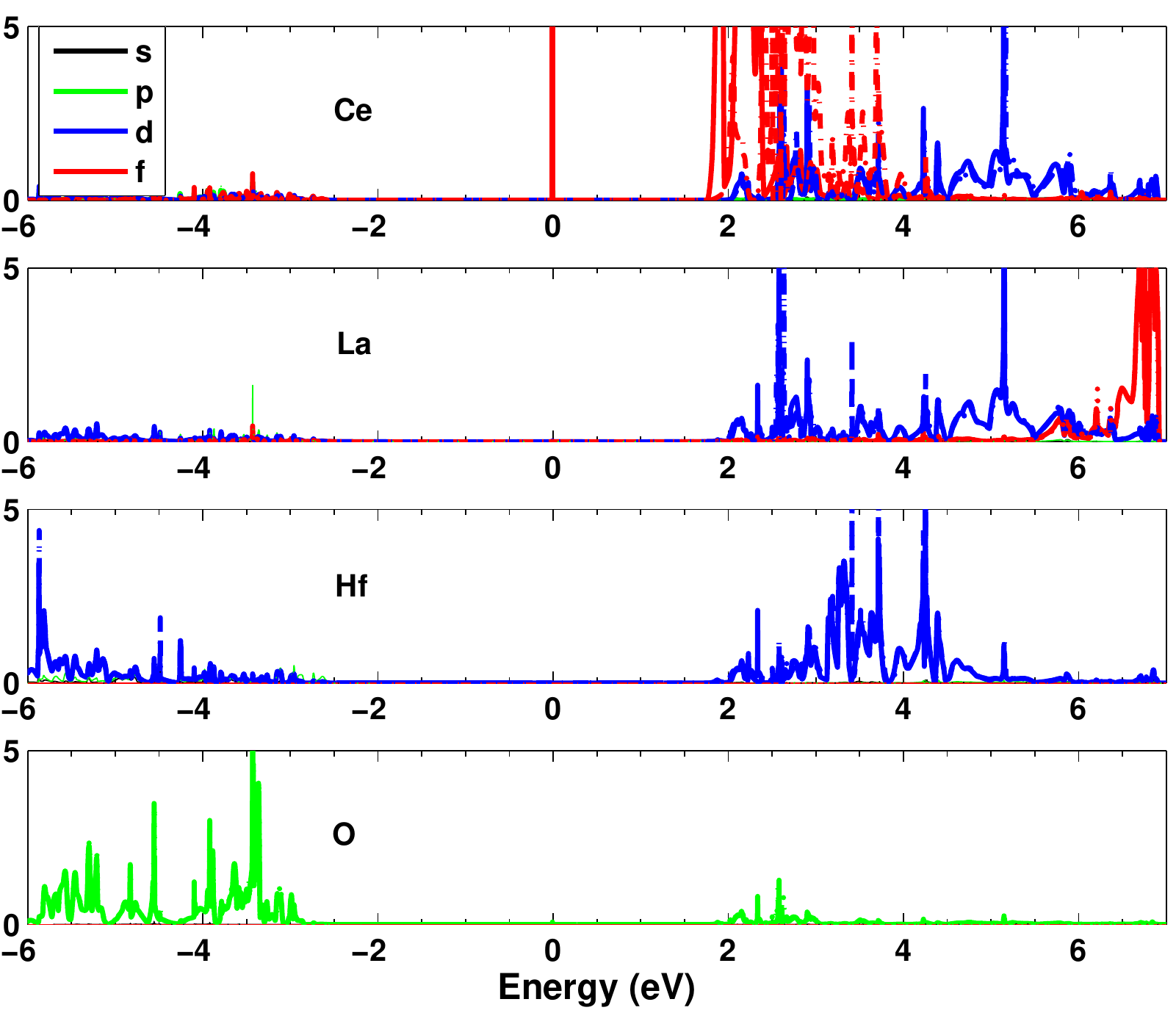}} \\   
\end{center}
\captionsetup{justification=justified}
\caption{\label{fig:fig4} Atom projected partial density of states (DOS) plots for Ce doped \subref{fig:4a} La$_2$Zr$_2$O$_7$ and \subref{fig:4b} La$_2$Hf$_2$O$_7$. Fermi level is set at 0eV. Spin up states are shown with solid lines while spin down with dashed lines. Oxygen 2p states form the top of valence band.}
\end{figure}

Figure~\ref{fig:fig4} shows the partial DOS plots for La$_2$Zr$_2$O$_7$:Ce and La$_2$Hf$_2$O$_7$:Ce. Oxygen 2\textit p states predominantly form the top of valence band. Because the host compounds have same cubic crystal structure, the Ce dopant environment is similar (surrounded by same number of nearest neighbor oxygen atoms) in the doped systems. The 4\textit f-VBM energy gap are therefore very similar for these systems. As we move from Zr to Hf compound, we see that the host conduction band states are progressively an admixture of La and transition metal \textit d states. This systematic trend of overlapping of transition metal \textit d states with La$^{3+}$ \textit d states has been analogously observed for SrMO$_3$ (M = Ti, Zr, Hf) oxides from spectroscopic measurements.\cite{Lee2010301} 


A common feature of these ground state plots is the positioning of unoccupied Ce 4\textit f states at the bottom of the conduction band. However, in a situation of hole capture by the Ce ion, Ce 4\textit f states will shift and move lower in energy. We have seen this movement of unoccupied Ce 4\textit f states (sometimes by about 2-3eV) from excited state calculations.\cite{canning:2010} This is due to the localized nature of the Ce 4\textit f states in these compounds. Considering the states above the Fermi level, for Ti and Zr compounds, Ce 5\textit d character states are located clearly above the bottom of conduction band. However, Ce 5\textit d states show mixing with the host \textit d character bands at the bottom of conduction band for La$_2$Hf$_2$O$_7$:Ce. This is noteworthy since the nature of conduction band states in this compound are very similar (5\textit d states of host ions La and Hf and Ce dopant) and we expect the DFT-PBE error in relative positioning of these states to cancel to a large extent. It is, therefore, important to check the localization of the excited state on the Ce site for La$_2$Hf$_2$O$_7$:Ce to estimate the possibility of a Ce 5\textit d-4\textit f transition.

\begin{figure}[h]
\begin{center}
\includegraphics[trim=0mm 0mm 0mm 0mm,clip,scale=0.5]{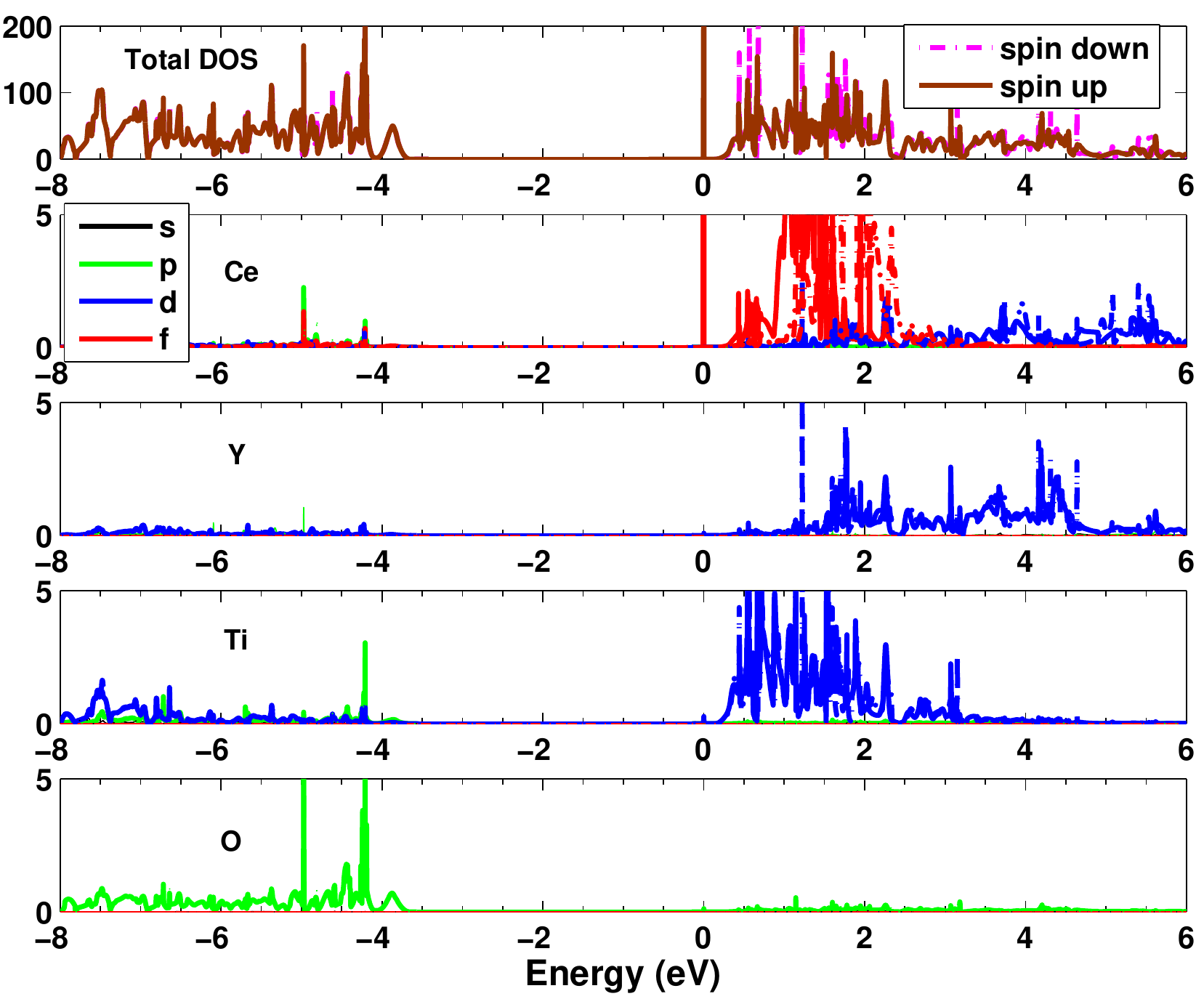}
\end{center}
\caption{\label{fig:fig5} Total and atom projected partial density of states plot for Y$_2$Ti$_2$O$_7$:Ce from PBE+U spin polarized calculations (U$_{\rm eff}$ = 2.5eV for Ce 4\textit f states and U$_{\rm eff}$ = 5.5eV for Ti 3\textit d states). Fermi level is set at 0. Ti 3\textit d states form the bottom of conduction band while Oxygen 2\textit p form the top of valence band.}
\end{figure}

We also studied Ce doped compounds of Y i.e., Y$_2$M$_2$O$_7$:Ce (M = Ti, Zr, Hf). Y$_2$Ti$_2$O$_7$ crystallizes in cubic pyrochlore structure however, Y$_2$Zr$_2$O$_7$ and Y$_2$Hf$_2$O$_7$ prefer the closely related disordered fluorite structure. The disordered fluorite structure exhibits the \textit{Fm3m} space group and has 
50\% A and B cation occupancy on the cation sublattice and a 7/8 occupancy 
on the 8c oxygen sublattice. This structure has the same stoichiometry as the pyrochlore structure, but is preferred when cation B is larger than cation A. Disordered fluorite structures of Y$_2$Zr$_2$O$_7$ and Y$_2$Hf$_2$O$_7$ are not listed in the ICSD so our calculations were limited to the related, but unstable, pyrochlore phase for these compounds. Figure~\ref{fig:fig5} shows the total and atom projected partial density of states plot for Y$_2$Ti$_2$O$_7$:Ce. Ground state DOS plots for this material show features similar to La$_2$Ti$_2$O$_7$:Ce with Ti 3\textit d states forming the bottom of the conduction band and O 2\textit p predominately at the top of valence band. Ce 5\textit d character states are situated much higher in the conduction band, which indicates that this material will not show Ce$^{3+}$ luminescence. Similar to La compounds, systematic shift in the position of transition metal \textit d states was observed in Y$_2$M$_2$O$_7$:Ce with Zr 4\textit d and Hf 5\textit d showing progressive overlapping with Y 4\textit d states. 


\subsection{\label{sec:ex}Excited State}

For most Ce-doped scintillator candidate materials, Ce 5\textit d states hybridize with the host conduction band states like La 5\textit d or Y 4\textit d states. A constrained LDA calculation gives us a qualitative measure of the position of the Ce 5\textit d state relative to the CBM and from our experience these calculations give consistent results for Ce doped scintillators and non-scintillators.\cite{canning:2010} However, for the class of the systems studied in this work, because the bottom of conduction band is mostly composed of the transition metal \textit d character states, it is expected that the localization would be mostly around the transition metal atoms.

\begin{figure*}
\subfloat[]{\label{fig:6a}\includegraphics[scale=0.37]{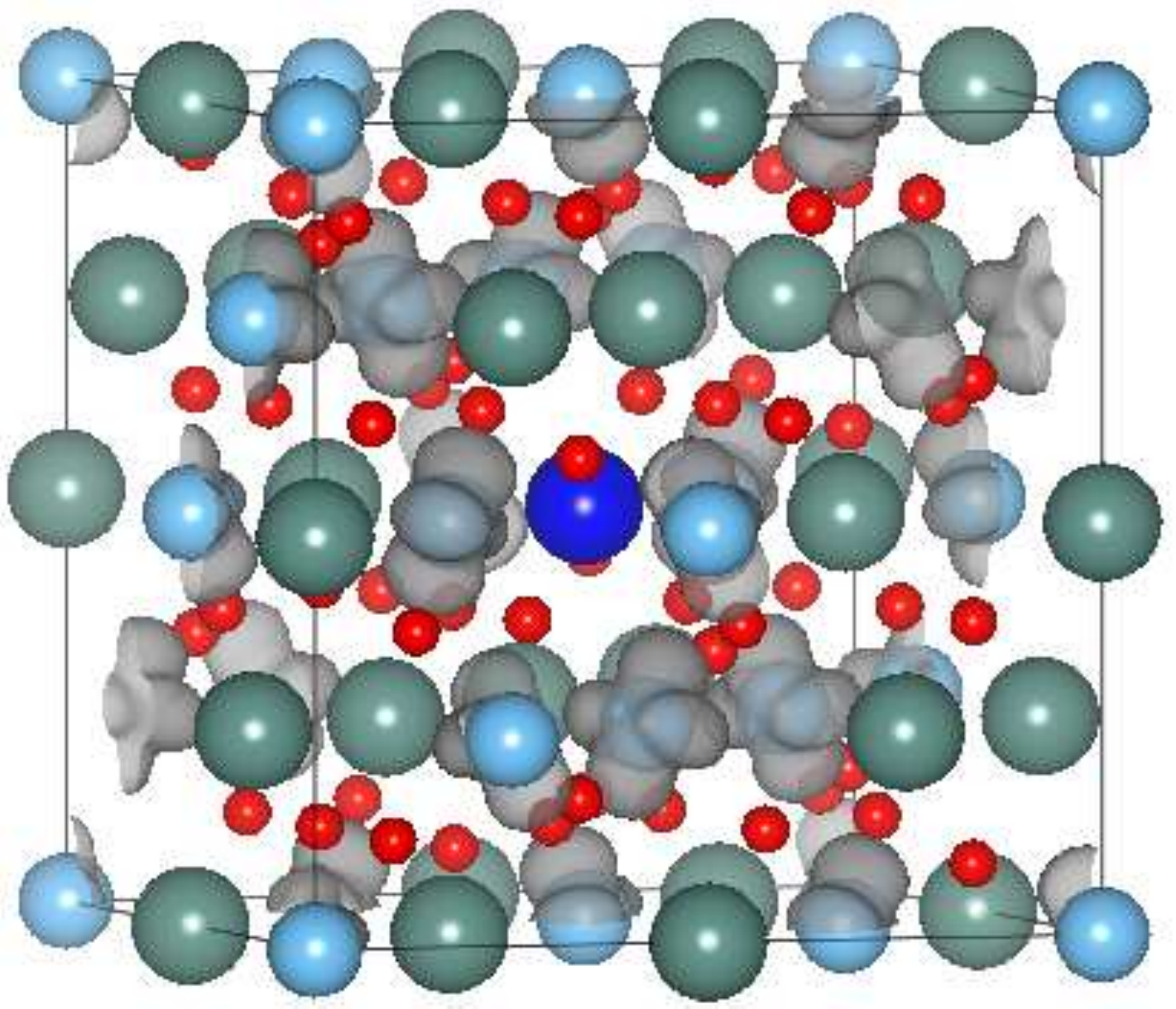}} \qquad
\subfloat[]{\label{fig:6b}\includegraphics[trim=0mm 0mm 0mm 0mm,clip,scale=0.37]{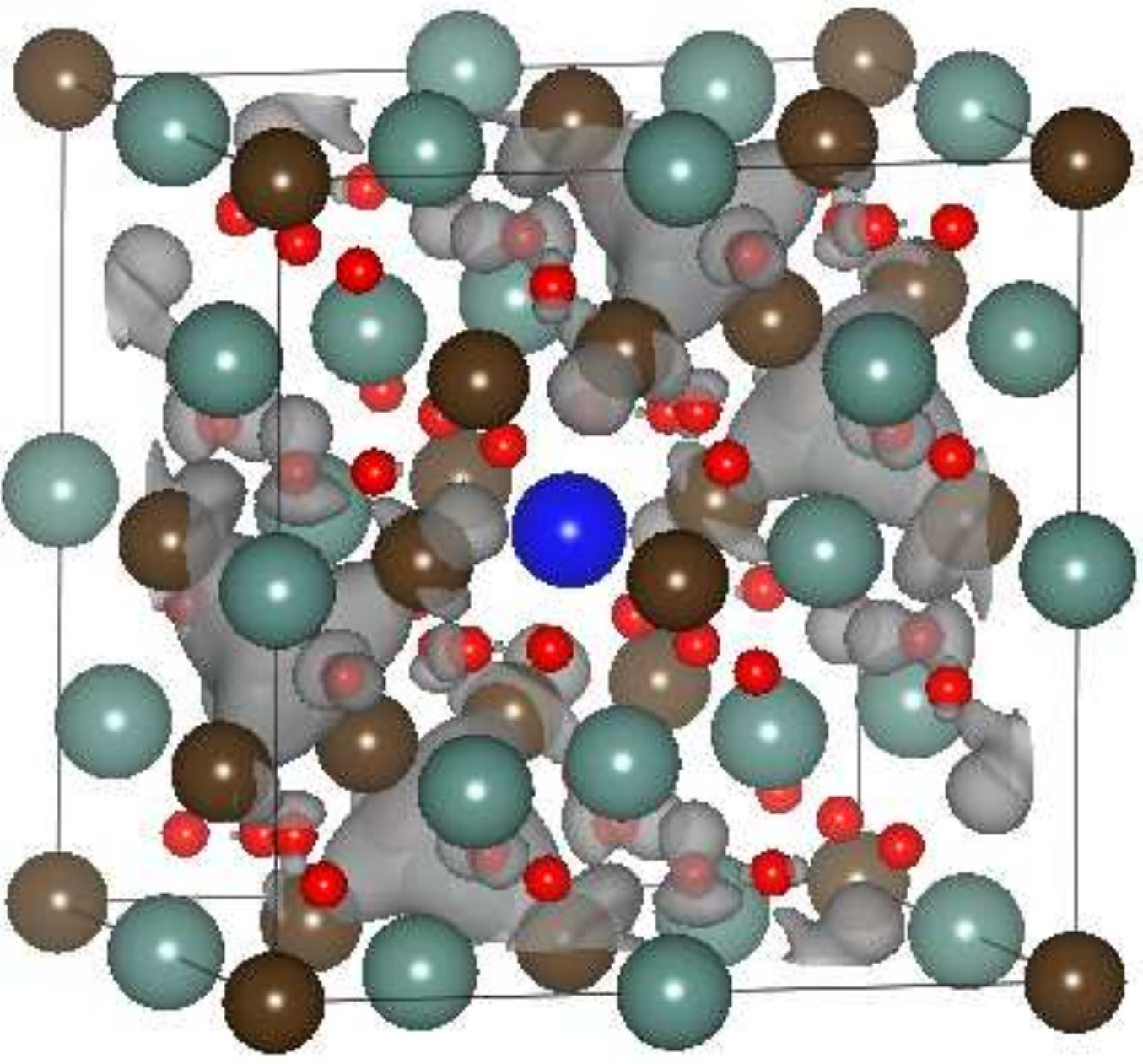}} \qquad   
\subfloat[]{\label{fig:6c}\includegraphics[scale=0.34]{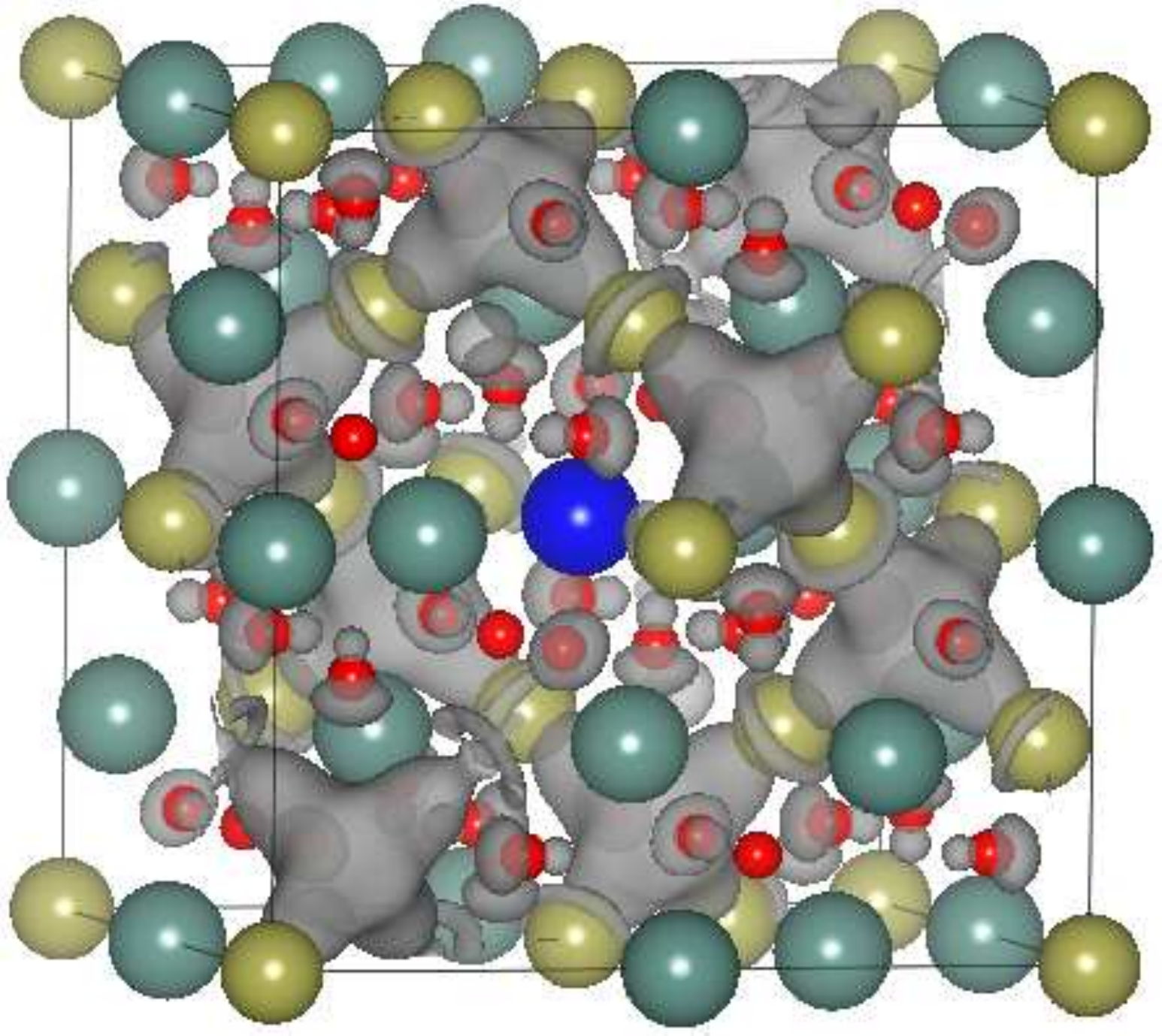}} \\
\subfloat[]{\label{fig:6d}\includegraphics[scale=0.38]{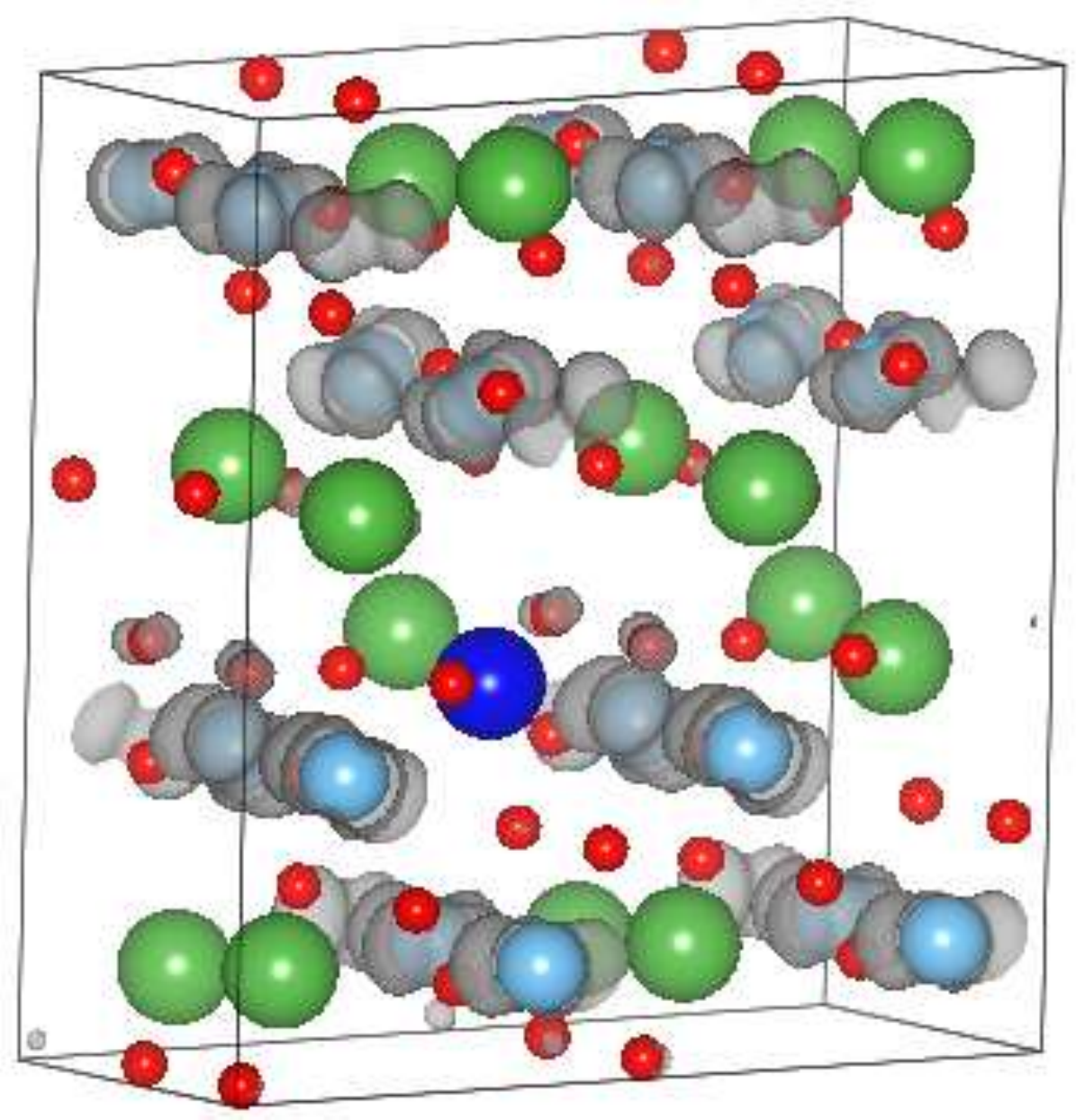}} \qquad
\subfloat[]{\label{fig:6e}\includegraphics[trim=0mm 0mm 0mm 0mm,clip,scale=0.39]{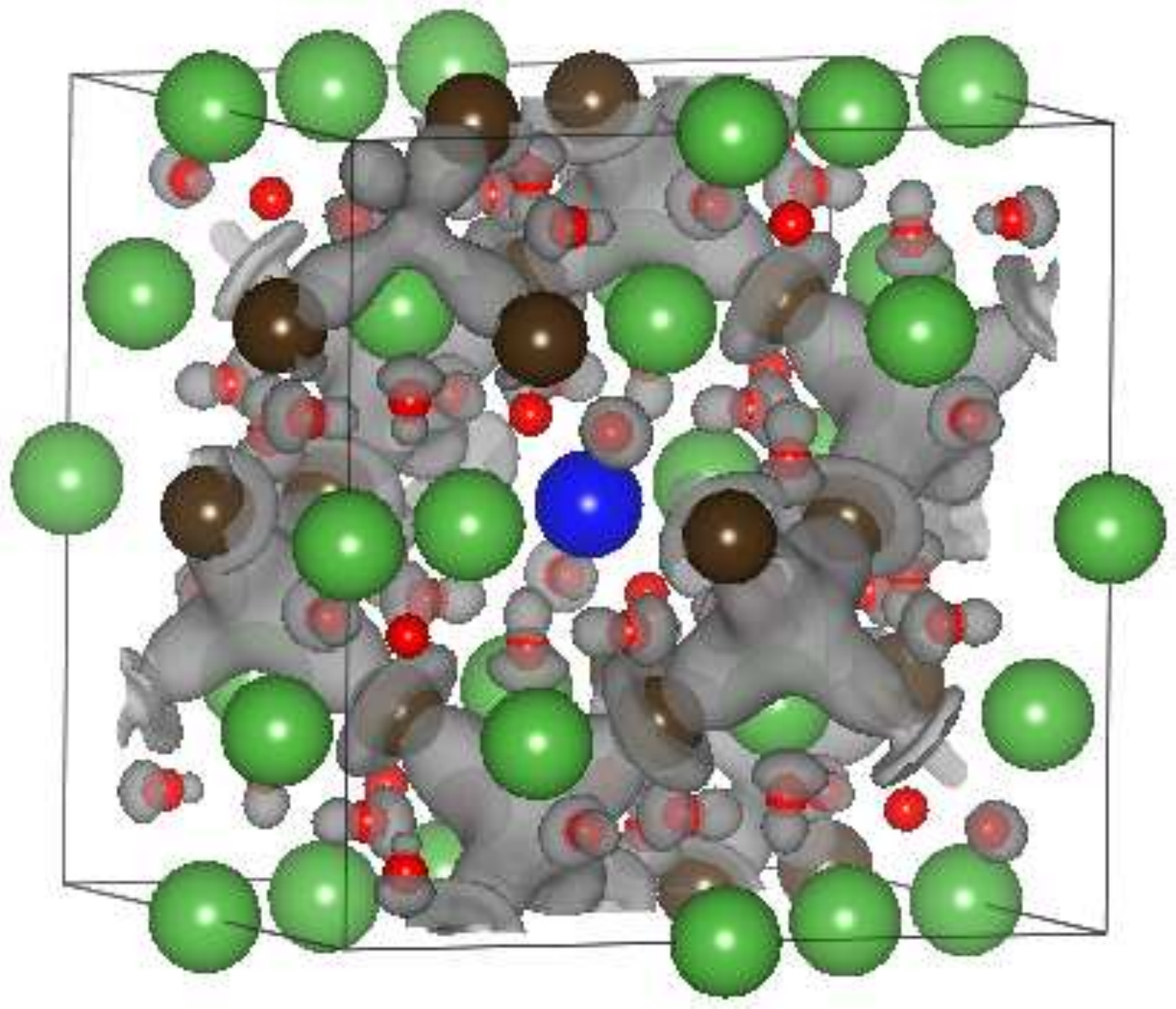}} \qquad
\subfloat[]{\label{fig:6f}\includegraphics[scale=0.33]{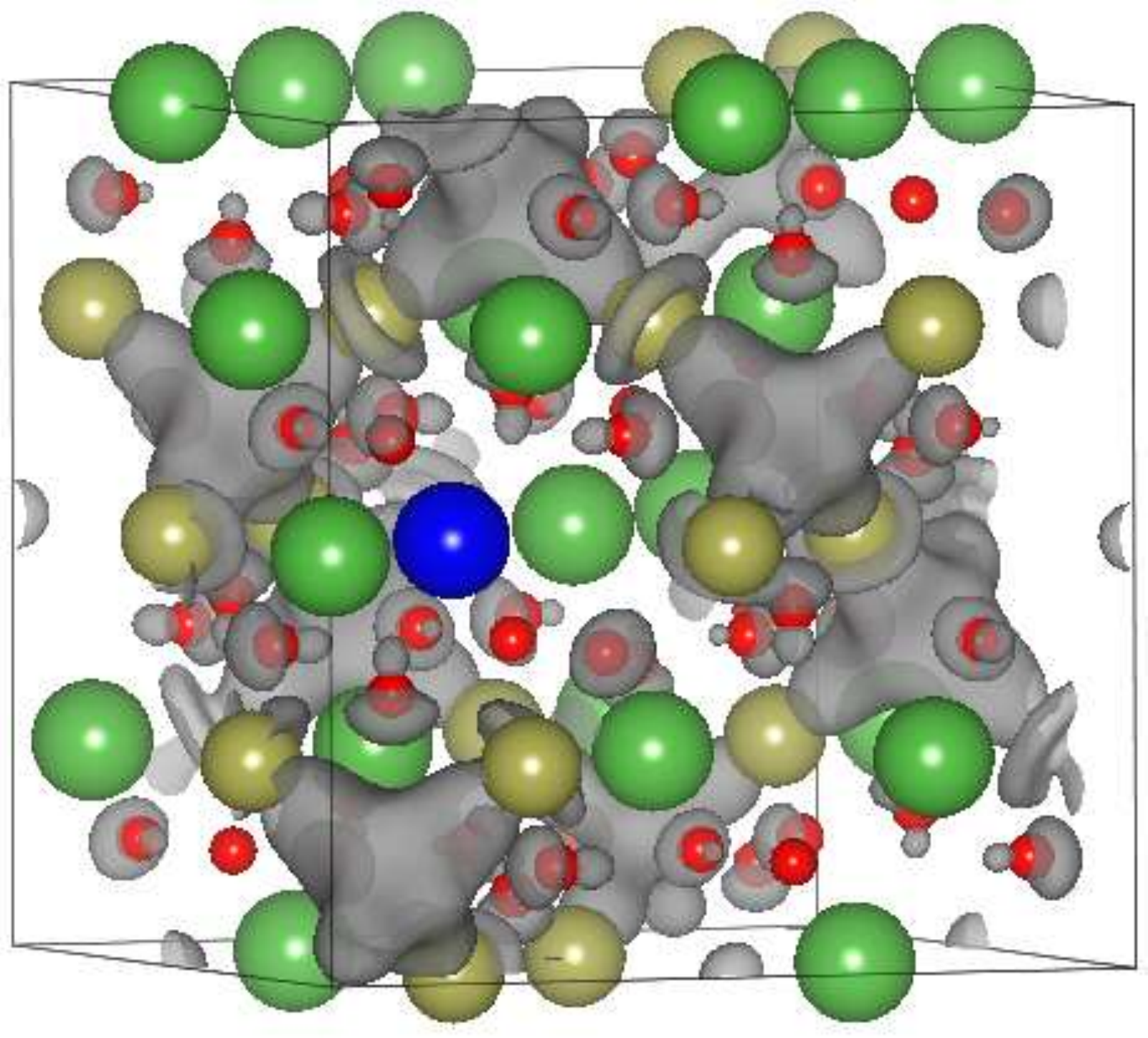}}\\
\includegraphics[scale=0.7]{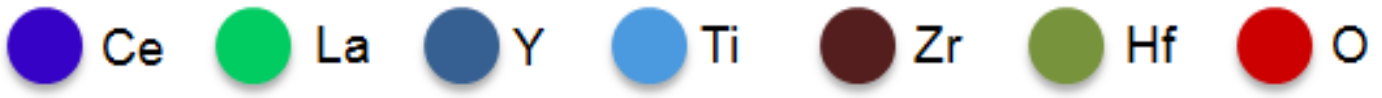}
\captionsetup{justification=justified}
\caption{\label{fig:fig6} Excited state plots for RE$_2$M$_2$O$_7$:Ce compounds (RE = Y, La; M = Ti, Zr, Hf). Plots show charge density isosurfaces of the lowest excited \textit d state at 50\% iso-surface threshold. \subref{fig:6a} Y$_2$Ti$_2$O$_7$; \subref{fig:6b} Y$_2$Zr$_2$O$_7$; \subref{fig:6c} Y$_2$Hf$_2$O$_7$; \subref{fig:6d} La$_2$Ti$_2$O$_7$; \subref{fig:6e} La$_2$Zr$_2$O$_7$; \subref{fig:6f} La$_2$Hf$_2$O$_7$. No localization on the Ce site indicates no Ce scintillation. Coloring scheme is indicated in the figure.}
\end{figure*}

Figure~\ref{fig:fig6} shows the charge density isosurface plots of the first \textit d character conduction band state for the compounds studied in this work. Localization on the Ce site would indicate possibility of Ce$^{3+}$ luminescence whereas a delocalized state indicates no possibility of Ce$^{3+}$ luminescence. We can clearly see that in all compounds studied there is no localization of the excited electron state on the Ce site. Therefore, our calculations indicate that the Ce 5\textit d states are situated much higher in energy in the conduction band and, as a consequence, ionized electron will not localize on the Ce site. Thus, Ce luminescence in these host materials is unlikely.


It is important to note that Ce$^{3+}$ activated AHfO$_3$ (A = Ba, Sr, or Ca) ceramic scintillators\cite{ji2005combustion,loef2007scintillation} have been reported in literature. These systems are however, quite different from the materials studied in this work. Foremost, there are no trivalent sites available for natural substitution of Ce$^{3+}$ (Ba, Sr and Ca are divalent while Hf, Zr and Ti are tetravalent) unlike La or Y compounds. Also, understanding the nature of charge compensation if, for example, Ce$^{3+}$ substitutes a Ba$^{2+}$, requires more work such as shown in reference.\cite{loureiro2005} Consequently, these hafnates are not subject of our present study.

During the course of our investigation we found more examples of La and Y compounds with Ti, Zr or Hf as constituents which can be easily identified as unsuitable candidates for Ce$^{3+}$ luminescence from first-principles calculations. Ba$_6$Y$_2$Ti$_4$O$_{17}$, LaLiTiO$_4$, LaNaTiO$_4$, YNaTiO$_4$, LaZrF$_7$, Y$_2$HfS$_5$ are examples of predicted non-scintillating Ce-doped host materials. 
To our knowledge there are no Ce$^{3+}$ luminescent La or Y host materials (phosphors or scintillators) with Ti$^{4+}$ and Zr$^{4+}$ as constituent ions, which agrees with our calculations. As far as La$_2$Hf$_2$O$_7$:Ce is concerned, our calculations agree with the observation of Ji \textit{et al.}\cite{ji2005part} and Cherepy \textit{et al.}\cite{cherepy2009scintillators} in that we believe Ce luminescence in this material is unlikely; although further experimental data on this material will be helpful. This material is the subject of experimental investigation by our group and detailed results will be published in the future. We would like to also mention that a recent experimental paper on Lu$_4$Hf$_3$O$_{12}$ powders doped with different rare-earth ions did not find any evidence of Ce activation in this host.\cite{havlak2010preparation} Havlak et al.\cite{havlak2010preparation} think that this is probably due to the positioning of lowest Ce 5d inside the host conduction band. This result is significant since Lu$_4$Hf$_3$O$_{12}$ crystallizes in the $\delta$-phase (space group R-3m) which is closely related to the fluorite structure. The pyrochlore and $\delta$-phase are oxygen deficient fluorite structure derivatives. Further information on these structures can be found, for example, in Ref.~[\onlinecite{sickafus2007radiation}].

Luminescence of Ti$^{4+}$ in La$_2$Hf$_2$O$_7$ host has been studied for application in X-ray imaging.\cite{ji20052} Our calculations for La$_2$Hf$_2$O$_7$:Ti clearly show the presence of Ti 3\textit d states well below the bottom of conduction band composed of an admixture of La and Hf 5\textit d states. This indicates that a charge transfer transition between the O 2\textit p states and Ti 3\textit d states is possible in agreement with measurements.

\begin{figure*}
\subfloat[]{\label{fig:7a}\includegraphics[trim=0mm 0mm 0mm 0mm,clip,scale=0.35]{lahfo_296.pdf}} \qquad
\subfloat[]{\label{fig:7b}\includegraphics[trim=0mm 0mm 0mm 0mm,clip,scale=0.35]{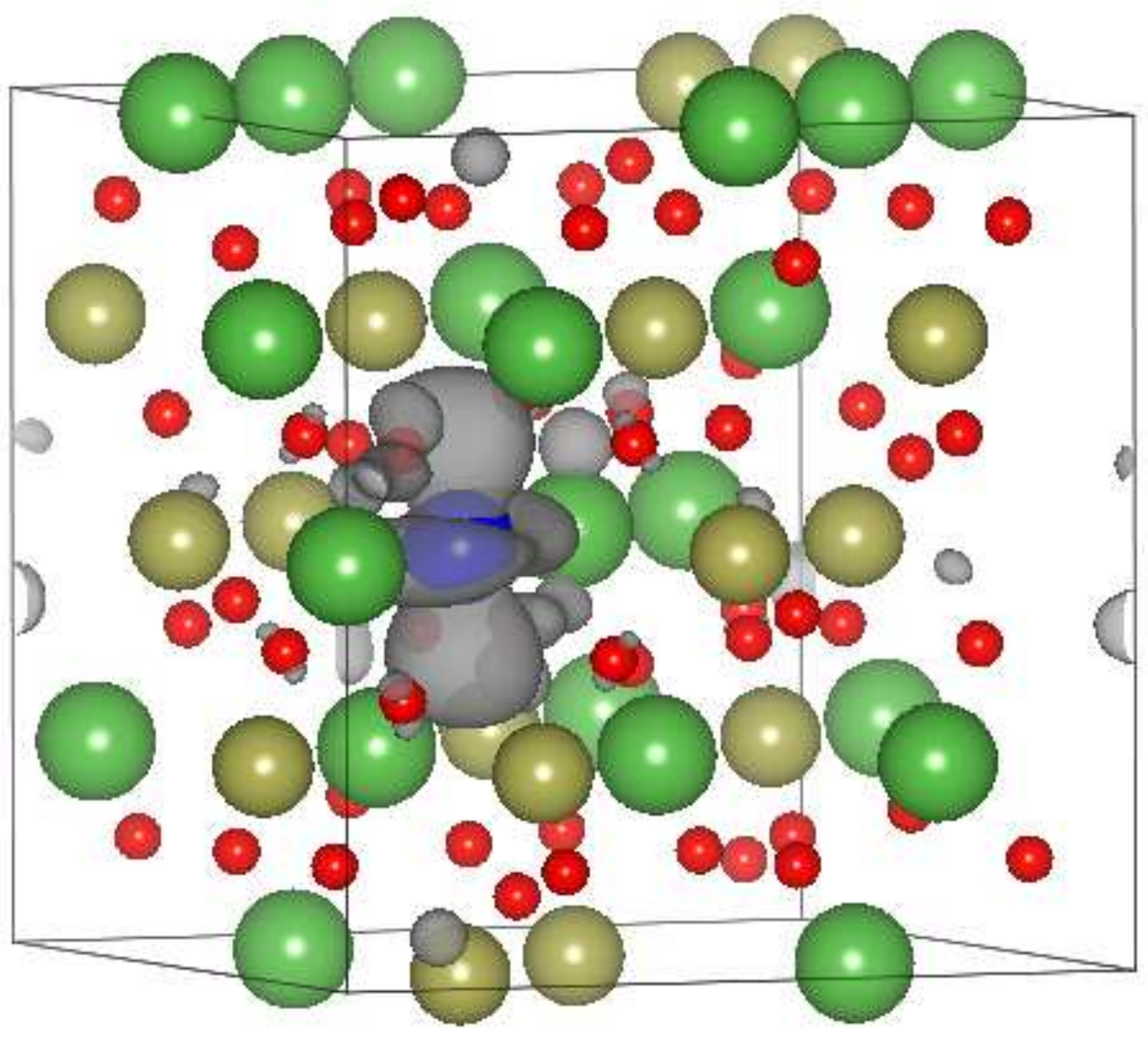}} \qquad   
\subfloat[]{\label{fig:7c}\includegraphics[trim=0mm 0mm 0mm 0mm,clip,scale=0.35]{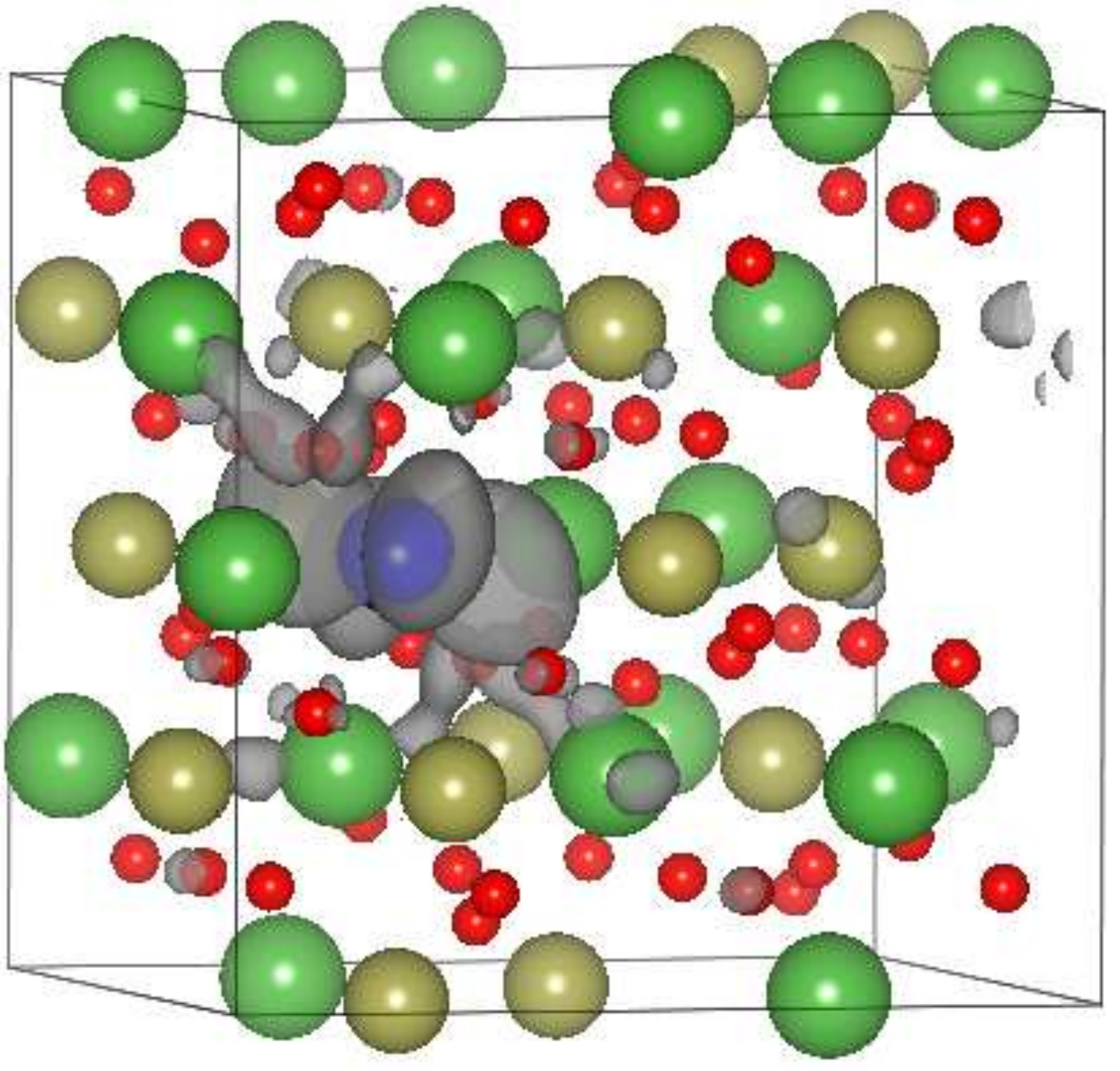}} \\
\captionsetup{justification=justified}
\caption{\label{fig:fig7} Charge density iso-surface plots of the three lowest excited states of La$_2$Hf$_2$O$_7$:Ce at the gamma point. \subref{fig:7a} lowest excited state (state \textit i) with mainly Hf 5\textit d character; \subref{fig:7b} next highest excited state (state \textit i+1) with mostly Ce 5\textit d character; \subref{fig:7c} next higher excited state (state \textit i+2) with predominately Ce 5\textit d character. State \textit i is separated from state (\textit i+1) by 40 meV while states (\textit i+1) and (\textit i+2) are separated by 6 meV.}
\end{figure*}

Figure~\ref{fig:fig7} further examines the three lowest conduction band states of La$_2$Hf$_2$O$_7$:Ce from excited state calculations. From our calculations we scan up the higher excited states and search for states which may have predominately Ce 5\textit d character. We find that, although the lowest excited state has a mainly Hf 5\textit d character, the next highest states ((\textit i+1) and (\textit i+2)) are mainly localized around the Ce site. This situation is analogous to the results presented by van der Kolk \textit{et al.}\cite{van2007luminescence} from experimental studies of LaAlO$_3$:Ce where they place lowest Ce 5\textit d states few tenths of an eV above the bottom of conduction band. Our calculations ignore the effects of spin-orbit coupling on the splitting of 5\textit d character states of the conduction band. A recent paper by Gracia et al.\cite{gracia2008ab} presented results from cluster calculations of Ce-doped Y$_3$Al$_5$O$_{12}$ (YAG). Ce$^{3+}$ substitutes for Y$^{3+}$ ion in an eightfold coordination of oxygens in YAG. Their results show that, upon including the effect of spin-orbit coupling the Ce 5\textit d states moved by about 0.1 eV higher in energy. Meanwhile, Garcia et al.\cite{garcia2004first} find that with the inclusion of spin-orbit interactions, Hf 5\textit d states shift by about 0.05 eV in HfO$_2$ using FLAPW calculations. This suggests that including the effects of spin-orbit coupling will not change the relative position of the lowest Ce 5\textit d state with respect to the bottom of the conduction band (Hf 5\textit d character).

Results presented so far have demonstrated one possible situation for no Ce luminescence wherein the lowest Ce 5\textit d is above the CBM (Figure 1(b)). This is also the reason why compounds such as LaAlO$_3$:Ce, La$_2$O$_3$:Ce, etc, do not show Ce$^{3+}$ luminescence, and our calculations show that in agreement with experimental results.\cite{canning:2010} In our studies of more than hundred Ce doped compounds so far we have not seen evidence of Ce 4\textit f situated in the host valence band of candidate scintillator materials. This is consistent with available experimental data on Ce doped compounds.

\section{Conclusion}

We have presented results of first-principles density functional theory calculations for Ce doped RE$_2$M$_2$O$_7$ compounds (RE = Y, La; M = Ti, Zr, Hf). Our calculations indicate that the lowest Ce 5\textit d state lies above the bottom of conduction band in these materials. This would prevent Ce luminescence in these materials. Our results agree with available experimental data. 

The predictive power of any pre-screen selection criteria is critically dependent on the successful prediction of negative candidates. This paper presents results of a class of materials which are found to be non-scintillators. These results indicate that our systematic calculation procedure can successfully screen non-scintillators and therefore, be applied to effectively identify candidate materials for Ce$^{3+}$ luminescence in inorganic compounds.

\begin{acknowledgments}
We would like to thank Y. Eagleman, E. Bourret-Courchesne and G. Bizarri for useful discussions during the course of this work. 
This work was supported by the U. S. Department of Homeland Security and carried out at the Lawrence Berkeley National Laboratory under U. S. Department of Energy Contract No. DE-AC02-05CH11231.
\end{acknowledgments}


%

\end{document}